\documentclass[journal ]{new-aiaa}
\usepackage[utf8]{inputenc}
\usepackage{textcomp}

\usepackage{graphicx}
\usepackage{amsmath}
\usepackage[version=4]{mhchem}
\usepackage{siunitx}
\usepackage{longtable,tabularx}
\title{Delaying transition induced by a strip of distributed roughness using additional fine grit roughness}

\author{Robin Joseph \footnote{} \footnote{PhD scholar, Department of Aerospace Engineering}}
\affil{Indian Institute of Science-Bangalore, India, 560022}
\author{P Phani Kumar\footnote{Scientist-D, Liquid Propulsion Systems Centre}}
\affil{Indian Space Research Organization, Thiruvananthapuram}
\author{Sourabh S Diwan\footnote{Assistant Professor, Department of Aerospace Engineering}}
\affil{Indian Institute of Science-Bangalore, India, 560022}

\begin{document}

\maketitle
\begin{abstract}
Distributed roughness occurs on aerodynamic surfaces like wind/gas turbine blades and aircraft wings causing early boundary layer transition resulting in higher skin friction, reduced lift-to-drag ratio, and lower power production. Despite being a recurring theme in practical engineering scenarios, the mechanism of boundary layer transition caused by distributed roughness is not well understood, and consequently, from a fluid dynamics perspective, methods for mitigating its effects are scarce. In this work, we present a passive method for delaying boundary layer transition caused by distributed roughness (simulated using a sandpaper strip) using a combination of secondary fine roughness strips placed immediately upstream and downstream of the distributed roughness. Hot-wire and PIV measurements are used to characterize the flow features and quantify transition delay. A combination of secondary roughness strips placed both upstream and downstream is shown to be most effective in delaying transition caused by the primary distributed roughness. Results suggest that the upstream roughness lifts the boundary layer reducing the effective Reynolds number of the primary roughness, while the downstream roughness reduces the strength of vortices shed from the primary roughness.  A parametric study on the length and type of secondary roughness shows that smooth strips can also delay the transition and there is an optimal length of the secondary roughness beyond which increasing the extent of the downstream roughness has marginal effects on transition delay. Analysis of the higher moments of fluctuating velocity shows that there are no specific signatures in the flow corresponding to the secondary roughness after the onset of transition which suggests that the secondary roughness can delay transition without substantially altering the transitional flow features. The results point to an adaptable and practical method for increasing the life cycle and efficiency of aerodynamic surfaces with distributed roughness while also generating insight into the mechanism of distributed roughness-induced transition. 
\end{abstract}


\section*{Nomenclature}
{\renewcommand\arraystretch{1.0}
\noindent\begin{longtable*}{@{}l @{\quad=\quad} l@{}}
U$_\infty$ & Free stream velocity\\

U$_{k}$      &     Mean velocity at height of the roughness\\
U$_{blasius}$    &      Normalized Blasius velocity \\
U   &   Mean velocity at a point\\

u$'$     &   Fluctuation velocity\\
u$_{rms}$      &    Root mean square of fluctuation velocity over the acquisition duration\\
u$_{max}$     &     Maximum value of u$_{rms}$  at a given streamwise location \\
\textit{k}       &      Height of roughness strip\\
\textit{f}     &     Frequency\\

$\delta^*$    &    Displacement thickness\\
$\delta$    &   99 \% Boundary layer thickness\\
\textit{H}    &     Shape factor\\

Re$_{k,crit}$     &     Critical Reynolds Number for distributed roughness\\
\textit{x},\textit{y},\textit{z}     &               Streamwise,wallnormal and spanwise coordinates\\
$\phi_{uu}$     &    Fourier amplitude of velocity\\
$\nu$     &      Kinematic viscosity\\
$\omega^{'}{z}$     &   Spanwise fluctuating vorticity\\
Re$_{\infty}$=U$_\infty$k/$\nu$    &     Reynolds number based on free stream velocity \\
$x_{t}$     &      Distance between the leading edge of the roughness and the transition onset location\\
Re$_{x}$=U$_\infty$\textit{x}/$\nu$    &     Streamwise Reynolds number \\
Re$_{Tr}$=U$_\infty$\textit{x}$_{t}$/$\nu$    &     Streamwise Reynolds number at the onset of transition \\
Re$_{k}$=k*U$_{k}$/$\nu$    &    Roughness Reynolds number based on roughness height and U$_{k}$ \\
St=\textit{f}*\textit{k}/U$_{k}$   &    Strouhal Number\\
U$_{d}$=U/{U$_\infty$}-U$_{blasius}$    &     Mean disturbance velocity\\

\end{longtable*}}

\section{Introduction}
During their operational life, aerodynamic devices like wind-turbine blades, gas-turbine blades, aircraft wings, etc. experience an accumulation of distributed roughness over their surfaces. Over wind-turbine blades and aircraft wings, roughness occurs due to insects, dust accumulation, and icing, whereas in gas-turbine blades, the surface roughness is typically due to the impact of ingested aerosols, accumulation of combustion products, corrosion, etc. The accumulated distributed roughness introduces disturbances into the flow causing the boundary layer over these surfaces to become transitional. For the distributed roughness typically found on aerodynamic devices, there is no specific order or pattern to the size and distribution of the roughness over the surface. Since distributed roughness frequently occurs in practical scenarios, it is of interest to understand the mechanism of distributed roughness-induced transition and develop reliable models for predicting transitional features. Studies (\cite{pinson2000effect}, \cite{suryanarayananmechanics}) have explored how the effect of distributed roughness can be mitigated and one expects that such methods will shed light on the dominant features and transition mechanisms in boundary layer flows over rough surfaces.  \par

Bons\cite{bons2010review} in a review on the effects of distributed roughness on gas turbine blades, described distributed roughness as causing early boundary layer transition, flow separation, and increased momentum thickness across the turbine blade resulting in increased specific fuel consumption and decreased efficiency. While roughness can often be characterized using centerline averaged roughness height measurements, significant difficulties exist in correlating this with the loss in power production/efficiency. A method often used is converting the distributed roughness into an equivalent sand grain roughness, suggested by Schlichting\cite{schlichting1979boundary}, but there are a vast variety of correlations that exist regarding this with engineers approaching each correlation on a case-by-case basis \cite{bons2010review}.  Several studies have worked on quantifying the losses due to roughness on gas turbine blades, for example, Bammert and Sandstede\cite{bammert1972measurements}, using sheets of sandpaper on single axis axial turbine demonstrated up to 19\% reduction in normalized efficiency. Severe cases of roughness accumulation would necessitate offline washing and engine overhaul of the turbine resulting in significant downtime.  \par

Wind turbines often develop distributed roughness due to insect strikes, sand erosion, and icing that can cause production losses of up to 25\%\cite{corten2001insects}.  Sandpaper strips have been used to demonstrate that distributed roughness reduces the lift-to-drag ratio\cite{pires2018experimental} of wind turbine blades. A study by Freudenreich et al.\cite{freudenreich2004reynolds}, using 60 grit carborundum roughness pasted on an airfoil demonstrated a 40\% reduction in lift-to-drag ratio. Studies have shown that dust accumulation mostly occurs near the leading edge\cite{khalfallah2007effect} where the turbine blade is most sensitive to roughness\cite{ren2009dust}.  The roughness causes early transition resulting in increased drag which is one of the reasons for underperformance. Similar to the case of gas turbine, the accumulated roughness often results in significant operational downtime and presents difficulties regarding cleaning due to the height at which wind turbines operate.\par  

There have been previous experimental attempts to understand the effect of distributed surface roughness on a boundary layer and most of these experiments use strips of sandpaper or sand grains to simulate distributed roughness. Using carborundum grits as roughness, Von Doenhoff and Horton\cite{von1956low} conducted measurements on the effect of distributed roughness on flow past an airfoil. They found that the presence of roughness causes early transition over the boundary layer and the streamwise extent of the roughness strips does not affect critical reynolds number. Levanthal and Reshotko\cite{reshotko1981preliminary} conducted measurements on flow over a flat plate with the entire surface covered with commercial sandpaper and found that the boundary layer undergoes transition at a much lower Reynolds number in comparison to the smooth plate. Langel et al.\cite{langel2014computational} and Kerho and Bragg \cite{kerho1997airfoil} used regular arrangements of molded hemispherical elements as distributed roughness on an airfoil and demonstrated early transition in comparison to the smooth case. Downs et al.\cite{downs2008transient} fabricated random distributed roughness using rapid prototyping and showed that disturbances generated by such roughness undergo transient growth. Recently, several studies have used strips of sandpaper to investigate the transitional flow downstream of a distributed roughness strip. Diwan and Morrison\cite{diwan2017spectral} used a strip of distributed roughness and freestream turbulence to generate a rapidly distorted boundary layer and demonstrated that the spectral structure resembled that of a turbulent boundary layer at moderately high Reynolds numbers. Anand and Diwan \cite{anand2020time} demonstrated that transition caused by a strip of sandpaper can be of two types: "spotty" or "non-spotty" based on the roughness Reynolds number, Re$_{k}$. Joseph and Diwan \cite{joseph2021growth} investigated the growth of disturbances in a pre-transitional boundary layer downstream of a sandpaper roughness strip and contrasted them with the pre-transitional region of FST-induced transition. Finally, Joseph et al. \cite{joseph2022characterization} used particle image velocimetry to visualize the mechanism of boundary layer transition caused by a sandpaper strip.\par

A relatively recent method of delaying transition caused by roughness is by using additional distributed roughness to manipulate the flow around roughness whose effect we desire to mitigate.  Previous studies have looked at delaying transition caused by isolated/array of roughness elements using a distributed roughness, commonly described as the " shielding effect". Kuester and White\cite{kuester2015roughness} experimentally demonstrated 'roughness shielding' by placing distributed roughness (generated by rapid prototyping) around a discrete roughness element and measuring the corresponding transition delay.  Suryanarayanan et al.\cite{suryanarayanan2020mechanisms} using naphthalene flow visualization demonstrated that a distributed roughness strip or a flat strip placed downstream of a discrete roughness element causes transition delay.  Moreover, using numerical simulations, they showed that transition delay is due to a reduction in streamwise vorticity generated from the discrete isolated roughness element by the distributed roughness placed downstream. Lu et al.\cite{lu2020investigation} reported simulations on delaying transition caused by isolated roughness by placing a two-dimensional strip downstream, wherein they demonstrated that a downstream roughness strip dissipates streamwise vorticity, delaying transition. They also reported that placing the two-dimensional strip immediately downstream of the discrete roughness element causes maximum transition delay and moving the two-dimensional strip further downstream has smaller effect on transition delay. They attribute this to the two-dimensional strip restricting the size of the separation zone behind the discrete roughness thereby reducing the initial disturbance amplitude. Suryanaraynan et al.\cite{suryanarayanan2020mechanisms} further show that an upstream distributed roughness delays transition caused by isolated roughness by lifting the boundary layer and reducing the effective Re$_k$  of the isolated roughness element. \par

Despite being a topic with extensive practical applications, relatively few studies have investigated methods for delaying boundary layer transition caused by distributed roughness. Pinson and Wang\cite{pinson2000effect} demonstrated that transition caused by distributed roughness can be delayed if the entire downstream region is covered with a finer grit roughness. For a 60-grit roughness placed near the leading edge, they demonstrated an increase in transitional Reynolds number by around 45\% when the entire downstream region was covered with 100-grit roughness. Using spectral analysis, they demonstrated that the step at the joint of the two roughness plays an important role in causing transition, and reducing this step size delays transition. More recently, using numerical simulations, Suryanarayanan et al.\cite{suryanarayananmechanics} demonstrated that boundary layer transition caused by a strip of sandpaper roughness can be delayed by placing a smooth strip of limited streamwise extent downstream of the roughness. However, in practical scenarios, it is likely that a smooth strip placed on an aerodynamic surface will become rough due to dust accumulation/icing. This is our motivation to investigate whether a fine roughness strip with limited streamwise extent can delay transition caused by a coarser distributed roughness. The requirement is that the additional roughness strip (secondary roughness) should be large enough to mitigate the effects of the primary roughness, but the disturbances caused by the roughness should not cause transition by themselves. \par

The objective of the present work is to demonstrate transition delay caused by strips of fine grit roughness (secondary roughness) placed adjacent to a coarse roughness (primary roughness). Transition delay is quantified by monitoring the increase in free stream velocity required for the flow to become transitional at a fixed streamwise location for each roughness configuration. Results show that while the downstream roughness strip delays transition, a combination of secondary roughness strips placed both upstream and downstream of the coarse roughness results in maximum transition delay. Further, consistent with \cite{suryanarayananmechanics}, the present work also demonstrates that a smooth strip (instead of a fine grit roughness) is also able to act as a secondary roughness delaying transition. The present study points to a lightweight, passive and inexpensive method for mitigating the reduction of efficiency due to naturally occurring distributed roughness on engineering devices. The downstream and upstream roughness strips delaying distributed roughness-induced transition also gives some insights into the mechanism of distributed roughness-induced transition.

\section{Experimental setup}

\begin{figure}[t]
\begin{center}
\begin{center}
    \includegraphics[width=.5\textwidth]{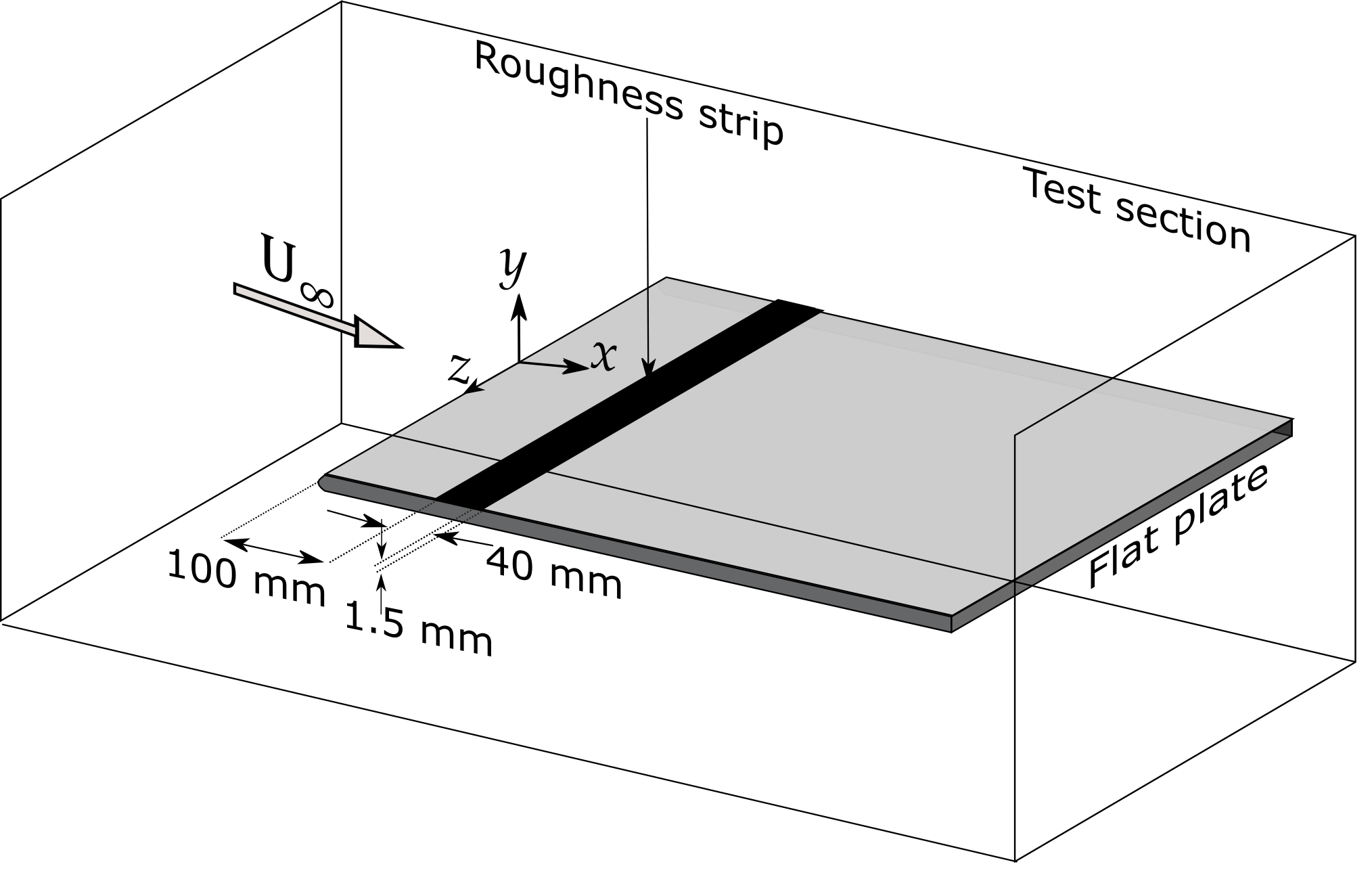}
    \end{center}
        \caption{Schematic of the flat plate experimental setup showing the single-strip configuration. The black strip denotes the 24-grit distributed roughness}
\label{Schematic_tunnel} 
\end{center}

\end{figure}

The experiments are conducted in low turbulence wind tunnels at the Department of Aerospace Engineering, Indian Institute of Science-Bangalore. The bulk of measurements are conducted on a 2.1 m long flat plate with super elliptic leading edge placed horizontally in the middle of the square test section with cross section dimensions 0.5 m × 0.5 m and length 3 m (Fig. \ref{Schematic_tunnel}). The free-stream turbulence intensity based on streamwise velocity fluctuation is 0.1\% at 10 m/s.  Zero pressure gradient is maintained over the flat plate and the variation of coefficient of pressure in the measurement region is within ±0.5\% \cite{anandthesis}. The non-uniformity in spanwise variation of the flow is estimated to be $\sim$1.09\% and rms variation is around 0.2\% \cite{anand2020time}. We also report limited measurements conducted on a 1.7m flat plate with a sharp leading edge placed inside another wind tunnel with a 0.6 m x 0.6 m cross-section and turbulence intensity of $\approx$ 0.1\%. The results reported in section III.D are from this wind tunnel.

\begin{figure}
\begin{center}
\begin{center}
    \includegraphics[width=0.8\textwidth]{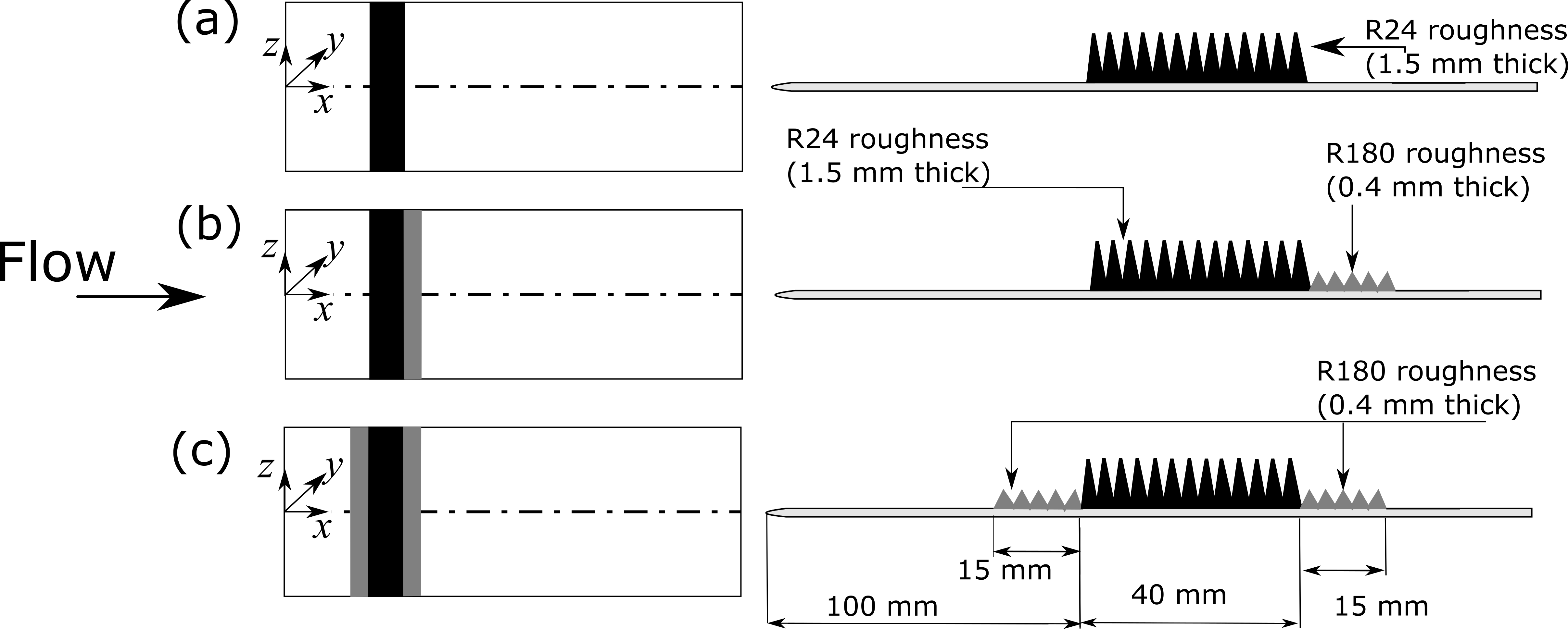}
    \end{center}
        \caption{Schematic of the three roughness configurations used (a) single-strip configuration, (b) two-strip configuration (c) three-strip configuration.}
\label{Schematic_configurations} 
\end{center}

\end{figure}

A schematic of the experimental setup showing the location of primary and secondary roughness strips in various configurations is shown in Fig. \ref{Schematic_configurations}. Here \textit{x}, \textit{y}, and \textit{z} denote streamwise, wall-normal, and spanwise coordinates, respectively. The primary roughness (R24 roughness) is a 24-grit extra coarse emery cloth of length (streamwise extent) 40 mm and thickness$\sim$1.5mm, while the secondary roughness (R180 roughness) is a 180-grit emery cloth of streamwise extent 15 mm and thickness$\sim$0.4 mm. The R24 roughness  consists of roughness grits (grit size 24, diameter $\approx$ 0.7mm) pasted on a backing cloth with the total thickness of the roughness strip(measured using Vernier scale) $\approx$ 1.5 mm. A detailed characterization of the R24 roughness, conducted using a laser scanner is provided in Joseph et al. \cite{joseph2022characterization}. Both R24 and R180 roughness spans the entire width of the flat plate and are pasted adjacent to each other without any gaps between them. Here, we conduct measurements for three configurations, namely the single-strip, two-strip, and three-strip configurations. The single-strip configuration (Fig. \ref{Schematic_configurations}a) contains only the primary roughness of 40 mm streamwise extent, whereas the two-strip (Fig. \ref{Schematic_configurations}b) and three-strip (Fig. \ref{Schematic_configurations}c) configurations contain both primary and secondary roughness strips. In the two-strip configuration, the secondary roughness (R180) with 15 mm streamwise extent is pasted downstream of the primary roughness (R24), while in the three-strip configuration, secondary roughness strips are pasted both upstream and downstream. The leading edge of the primary roughness is located at a distance of 100 mm from the plate leading edge (Fig. \ref{Schematic_configurations}).\par

Hot-wire measurements are made along the plate centerline using the Dantec Dynamics single axis constant temperature anemometer (CTA) system (Streamline pro). Data acquisition is carried out using a National Instruments DAQx board and LabView software. Hot-wire data is acquired for 60 seconds at a sampling rate of 20 kHz and low pass filtered at 10 kHz to remove aliasing error following the Nyquist criterion. The hot-wire probe (55P11) is calibrated by placing it beside a Pitot-static tube in the free-stream and by using King’s law E$^{2}$=A+B$\sqrt{U}$, where E is the Hot-wire voltage, U is the velocity measured using the Pitot-static tube and A \& B are calibration constants). The differential pressure from the pitot-static tube is measured using a digital manometer (Furness Controls, FCO560). The uncertainty in the measured mean velocity (U) and root mean square of fluctuation velocity (u$_{rms}$) is estimated to be 0.9\%. Further details on the setup and uncertainty calculations can be found in\cite{anand2020time}.\par

Limited Particle image velocimetry (PIV) measurements are conducted downstream of the single-strip configuration to characterize the flow features. The PIV setup consists of a double cavity Nd:YAG laser(100 mJ/pulse), a Sharpvision 1400 DE CCD camera (1360 x 1036 pixel resolution) with a 50 mm focal length lens, and the proVISION processing software.  Wall-normal (\textit{x}-\textit{y}) plane measurements are conducted along the centerline downstream of the roughness and spanwise (\textit{x}-\textit{z}) plane measurements are taken at a height of \textit{y}$\sim$1mm from the wall. For more details on the PIV setup and data processing, see\cite{kumar2015effect}.

\begin{figure}[t]
\begin{center}
\begin{center}
    \includegraphics[width=1\textwidth]{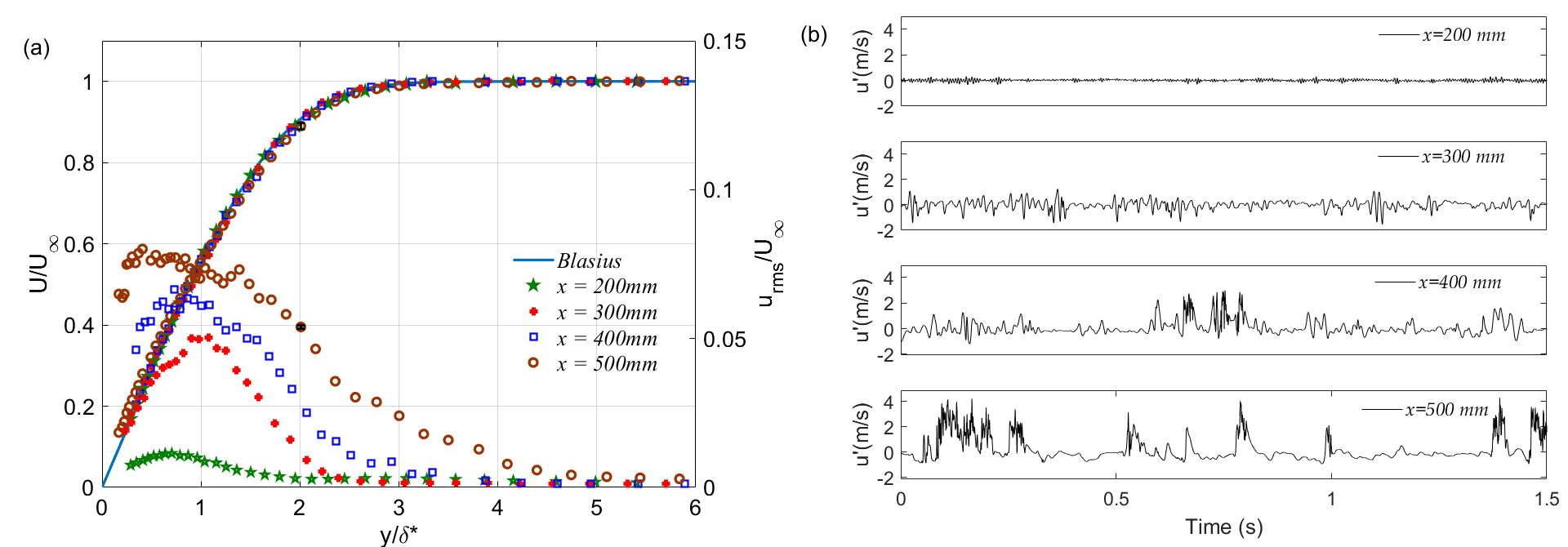}
    \end{center}
        \caption{Flow characterization of single-strip configuration with U$_{\infty}$ = 7.44 m/s. (a) Mean and RMS velocity profiles compared with Blasius profile; $\delta*$ is the boundary layer displacement thickness. (b) Fluctuating velocity signals at different streamwise locations}
        \label{U_avg}
\end{center}
\end{figure}

\begin{figure}[t]
\begin{center}
\begin{center}
    \includegraphics[width=1\textwidth]{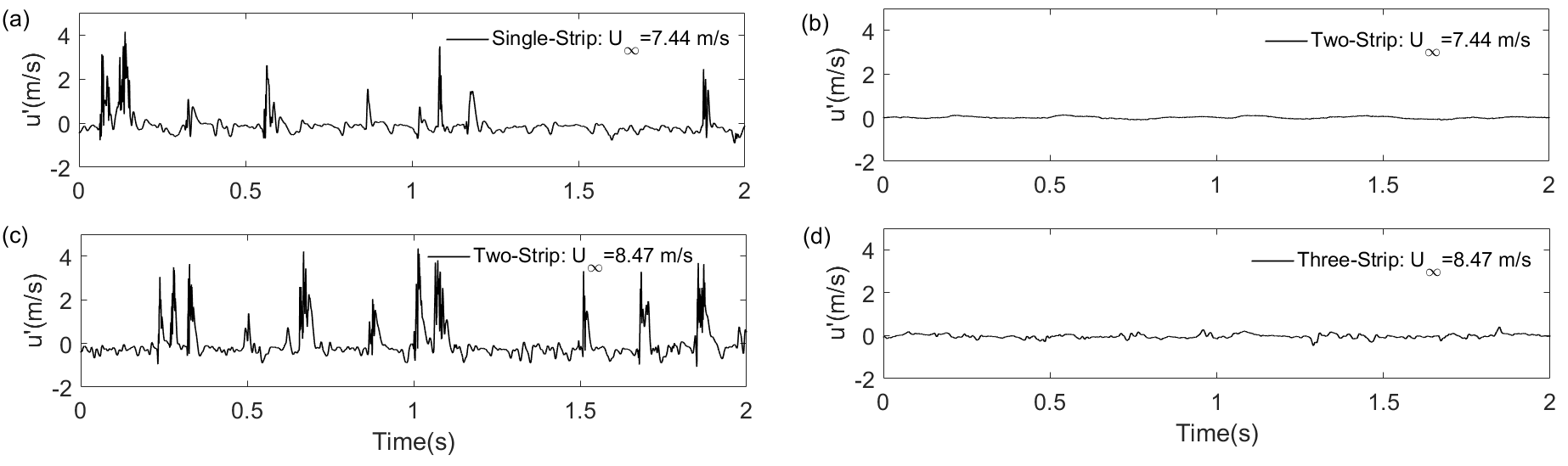}
    \end{center}
        \caption{Fluctuating velocity signals at \textit{x}=500 mm for different configurations (a) Single-strip configuration U$_\infty$ = 7.44 m/s (b) Two-strip configuration U$_\infty$ = 7.44 m/s (c) Two-strip configuration U$_\infty$ = 8.47 m/s (d) Three-strip configuration U$_\infty$ = 8.47 m/s }
        \label{u_inst}
\end{center}
\end{figure}
\section{Results and discussion}
\subsection{Transition delay by addition of secondary roughness}

First, we characterize the transitional flow downstream of the single-strip configuration. While there are several definitions on how to define the start of the transition zone, intermittency reaching $\approx$ 10\% is a definition that has been used before\cite{fransson2005transition} and is used in the present work for the first set of measurements. The intermittency of a velocity signal is defined as the fraction of time for which the signal is turbulent and is calculated using the method in Fransson et al. \cite{fransson2005transition}. We choose \textit{x} = 500 mm as a convenient location for transition onset and adjust the free stream velocity (U$_\infty$) to achieve an intermittency value of $\approx$ 10\% at this location. For the single-strip configuration, a free stream velocity of U$_\infty$ = 7.44 m/s is required for the flow downstream of the boundary layer to become transitional at \textit{x} = 500 mm.  For the single-strip configuration at U$_\infty$=7.44 m/s, Re$_k$ =551 (Re$_{k}$=k*U$_{k}$/$\nu$), which is lower than the critical Reynolds number, Re$_{k,crit}$ = 600 generally observed for distributed roughness\cite{langel2014computational} and hence the boundary layer is pre-transitional and the transition onset is nearly 400 mm downstream of the leading edge of the roughness.\par

The mean velocity profiles at different streamwise locations downstream of the random primary roughness are plotted in Fig. \ref{Schematic_configurations}a along with the numerically computed Blasius profile. The mean velocity profiles closely follow the Blasius profile, except at \textit{x} = 500 mm, where the mean velocity profile is slightly fuller with an accompanying reduction in the shape factor (\textit{H} = 2.38), which is expected as the flow becomes transitional there. Fig. \ref{Schematic_configurations}a also plots the root mean square fluctuation velocity profiles (u$_{rms}$) for the same streamwise locations. At all downstream locations, the peak of u$_{rms}$ profiles is at \textit{y}/$\delta^*$ $\leq$1 with the peak moving closer to the wall as the flow becomes transitional.
The small scatter in the u$_{rms}$ profiles is generally observed in the early stages of transition and does not imply measurement inaccuracies. This has been confirmed by carrying out measurements for acquisition durations higher than 60s which also showed a similar scatter. Note that the acquisition time of 60 seconds corresponds to a boundary layer turnover time (\textit{tU$_\infty$/$\delta$}; where t is the acquisition duration, and $\delta$ is the 99\% boundary layer thickness) of 72941 at \textit{x} = 500 mm for the single-strip configuration, which is large enough to obtain good velocity statistics \cite{bradshaw2013introduction}. Note that all the other measurements presented in this work have higher boundary layer turnover time than 72941. Repeatability measurements were also carried out which did not change the nature of u$_{rms}$ profiles. It is worth noting that a scatter in u$_{rms}$ profiles at low intermittencies was also observed by Matsubara et al.\cite{matsubara2001disturbance} and Joseph and Diwan \cite{joseph2021growth}. Figure~\ref{Schematic_configurations}b plots the fluctuating velocity signals at different streamwise locations downstream. At each streamwise location, the velocity signal corresponding to maximum u$_{rms}$ (i.e. u$_{max}$) is plotted. The gradual change from laminar flow at \textit{x} = 200 mm to transitional flow at \textit{x} = 500 mm is clearly seen with the appearance of turbulent spots in the velocity signal at \textit{x} = 400 mm $\&$ 500 mm; the latter being the 10\% intermittency location.   \par

\begin{figure}[t]
\begin{center}
\begin{center}
    \includegraphics[width=1\textwidth]{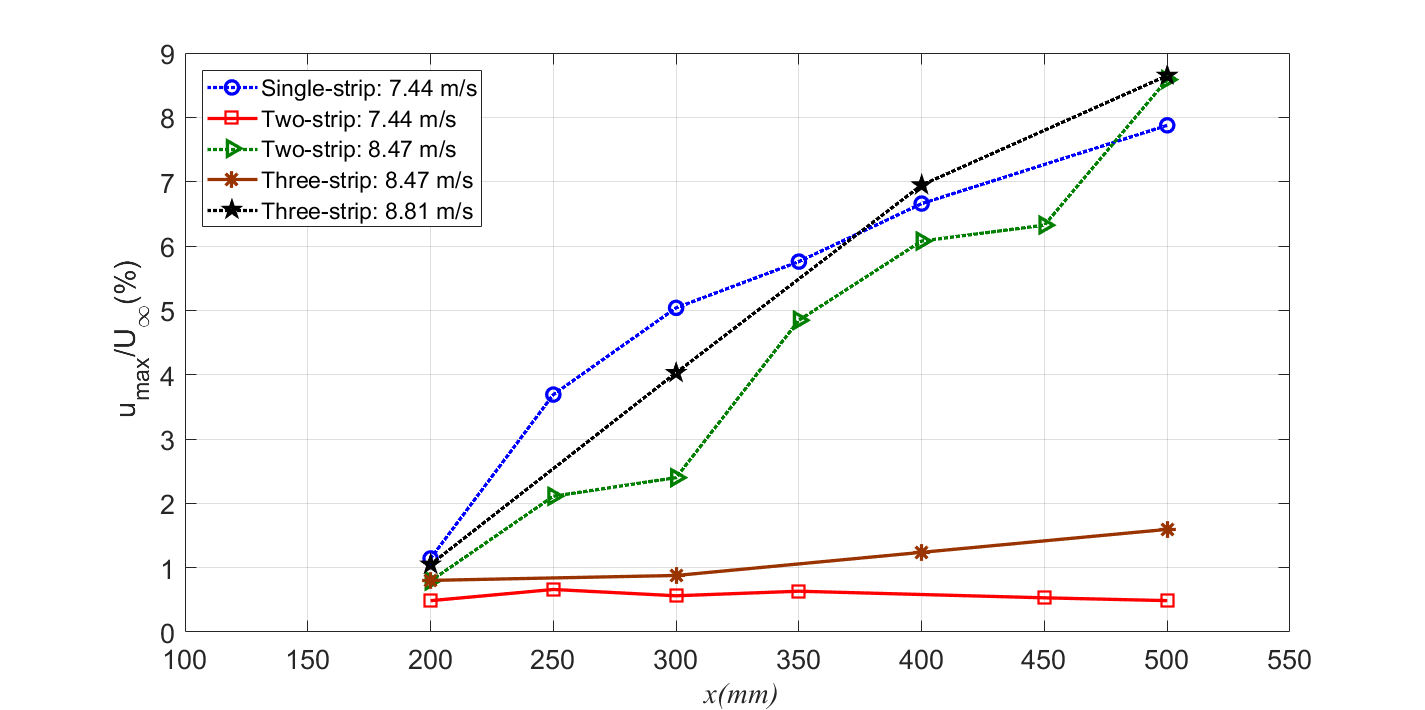}
    \end{center}
        \caption{Development of u$_{max}$ for different roughness configurations at different free-stream velocities}
\label{u_max} 
\end{center}
\end{figure}

Next, we study the effect of addition of secondary roughness strips on the transition onset location. Figure \ref{u_inst} plots the fluctuating velocity signals at \textit{x} = 500 mm for different roughness configurations and Fig. \ref{u_max} plots the u$_{max}$/U$_\infty$ (u$_{max}$ is the maximum u$_{rms}$ at a given streamwise location) with the streamwise distance, \textit{x}. Comparing Fig. \ref{u_inst}a  and \ref{u_inst}b, we see that with the addition of the secondary roughness strip downstream of the primary roughness, the fluctuating velocity signal becomes laminar (turbulent spots are not seen) at U$_\infty$=7.44m/s. We also observe a corresponding decrease in u$_{max}$/U$_\infty$ (u$_{max}$ is the maximum value of u$_{rms}$ at a given streamwise location) from 8\% (single-strip configuration;U$_\infty$=7.44 m/s) to less than 1\% (two-strip configuration;U$_\infty$=7.44 m/s) as seen in Fig. \ref{u_max}. To determine at what speed transition happens for the two-strip configuration, we increase the free stream velocity until the measured intermittency at \textit{x} = 500 mm is $\sim$10\% and find it to be U$_\infty$ = 8.47 m/s (13\% higher than the single-strip configuration). This confirms that the addition of the secondary roughness strip downstream delays transition caused by the primary roughness.\par

The transition delay obtained by adding a secondary roughness could be attributed to  the reduction in the severity of backward-facing step present at the trailing edge of the primary roughness\cite{pinson2000effect}. To explore this idea further, we use the three-strip configuration in which another R180 strip is pasted immediately upstream of the primary roughness to reduce the severity of the forward step. Figure. \ref{u_inst}c and Fig. \ref{u_inst}d plots the fluctuating velocity signals for two-strip and three-strip configuration at U$_\infty$=8.47 m/s and we see that the flow is laminar in the three-strip configuration. Correspondingly, in Fig. \ref{u_max}, there is a decrease in  u$_{max}$/U$_\infty$ at \textit{x} = 500 mm from 8.5\% (two-strip configuration; U$_\infty$=8.47 m/s) to 1.5\% (three-strip configuration;U$_\infty$=8.47 m/s). This confirms that the addition of secondary roughness upstream of the primary roughness further delays transition. Finally, the free stream velocity has be increased to U$_\infty$=8.81 m/s (18\% higher than the single-strip configuration) for the three-strip configuration to become transitional (fig \ref{u_max}). The results are summarized in Tables \ref{table1} and \ref{table2}.

\begin{table}[h!]
\begin{center}
\begin{tabular}{ |c|c|c|c|c| } 
 \hline
 Roughness configuration & U$_{\infty}$=7.44 m/s & U$_{\infty}$=8.47 m/s & U$_{\infty}$=8.81 m/s\\ 
  \hline
 Single-strip: \textit{x} = 500 mm & \textbf{Transitional} & \textbf{Transitional} & \textbf{Transitional} \\ 
  \hline
 Two-strip: \textit{x} = 500 mm & Laminar & \textbf{Transitional} & \textbf{Transitional}\\ 
  \hline
  Three-strip: \textit{x} = 500 mm & Laminar & Laminar & \textbf{Transitional}\\ 
    \hline

\end{tabular}
\caption{ Free stream velocity required for different roughness configurations to become transitional at \textit{x} = 500 mm.}\label{table1}
\end{center}

\end{table}

\begin{table}[h!]
\begin{center}
\begin{tabular}{ |c|c|c| } 
 \hline
 Roughness configuration & U$_{\infty}$ & Re$_{x}$(Transitional)\\ 
  \hline
 single-strip: \textit{x} = 500 mm & U$_{\infty}$=7.44m/s & 2.13x10$^5$\\ 
  \hline
  two-strip: \textit{x} = 500 mm & U$_{\infty}$=8.47m/s & 2.43x10$^5$\\ 
  \hline
 three -strip: \textit{x} = 500 mm & U$_{\infty}$=8.81m/s & 2.53x10$^5$\\ 
    \hline

\end{tabular}
\caption{ Parameters corresponding to transition onset at \textit{x} = 500 mm.}\label{table2}
\end{center}

\end{table}

\subsection{Spectral characteristics of different roughness configurations undergoing transition}
To gain insight into the mechanism of transition delay caused by the addition of secondary roughness, spectral analysis of fluctuating velocity signals for different roughness configurations is carried out in the following sections. PIV measurements made for the single-strip configuration are also used here to visualize flow features. The pre-multiplied spectral amplitude is normalized using \textit{u}$_{max}$ to quantify the relative contribution of frequencies to the total fluctuating energy. The spectral frequency is normalized using \textit{k}=1.5 mm, height of the R24 roughness and U$_k$=6.41  m/s, which is the velocity at the height of the roughness for the single-strip configuration, obtained from a Blasius profile. All the spectra plotted in this work have frequency non-dimensionalized using the aforementioned \textit{k} and U$_k$ values. 

\subsubsection{Single-strip configuration}
The roughness used in the single-strip configuration is the 24-grit extra coarse emery cloth roughness. Due to the randomly distributed nature of the sandpaper roughness grits, one can expect it to induce broadband disturbances downstream\cite{diwan2017spectral}. Further, due to the backing cloth, the roughness strip can be expected to act as a forward-backward step that sheds spanwise vortices downstream. Finally, the distributed roughness has also been demonstrated to create streamwise vortices downstream (\cite{drews2012direct}, \cite{suryanarayananmechanics}). Note that the streamwise vortices generated by isolated/array of roughness elements are generally steady/quasi-steady\cite{fransson2004experimental}.

\begin{figure}
\begin{center}
\begin{center}
    \includegraphics[width=1\textwidth]{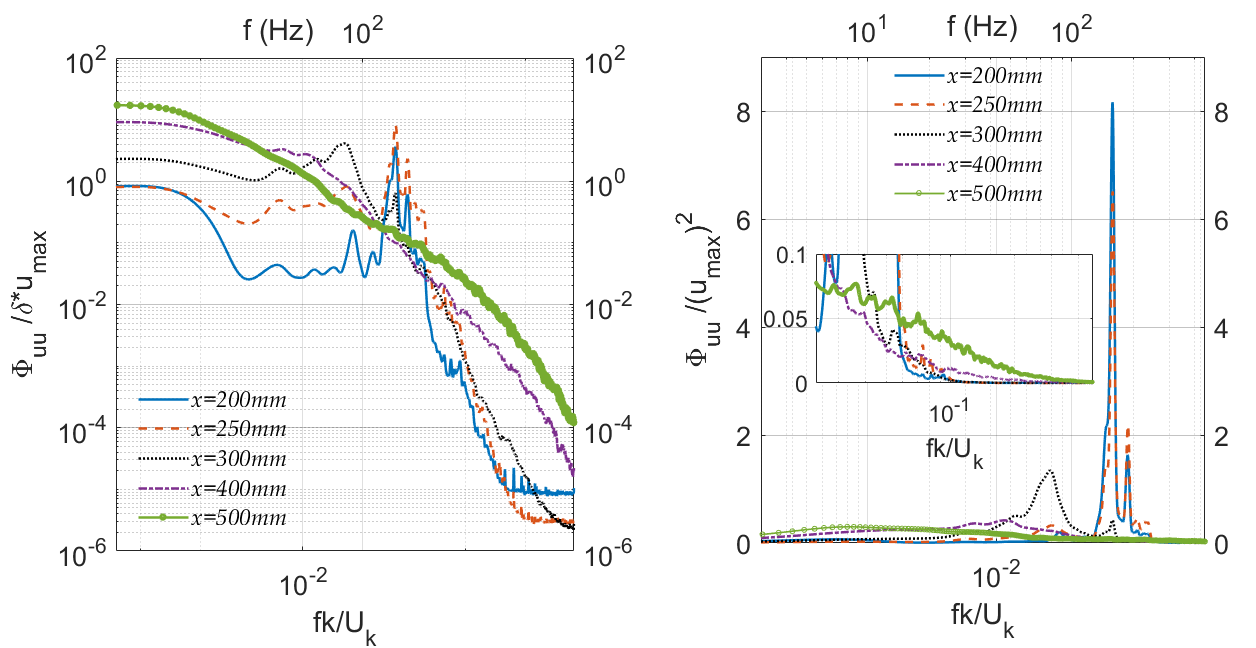}
    \end{center}
        \caption{(a) Power spectral density and (b) pre-multiplied spectra at different streamwise locations for single-strip configuration at U$_{\infty}$=7.44 m/s.}
\label{Single_strip_spectra} 
\end{center}
\end{figure}

Figure \ref{Single_strip_spectra}a and Fig. \ref{Single_strip_spectra}b  plot the power spectral density and pre-multiplied spectra for the single-strip-configuration at different streamwise locations corresponding to a free-stream velocity of 7.44 m/s. The power spectral density plotted here is the spectral power ($\phi_{uu}$) normalized by \textit{u}$_{max}$ and displacement thickness ($\delta^*$). At \textit{x} = 200 mm, the closest streamwise location to the roughness where measurements are taken, there is a peak in the spectrum at 158 Hz, corresponding to a Strouhal number (f\textit{k}/U$_{k}$) of 0.037. A similar value of Strouhal number (f\textit{k}/U$_{k}$) of 0.032 was reported in hotwire measurements conducted downstream of the same grit roughness at U$_{\infty}$=7.0 m/s in our results presented in Joseph et al.\cite{joseph2022characterization}. We attributed the peak in the spectrum to vortices shed from the forward-backward step caused by the roughness strip consistent with previous studies such as Pinson and Wang\cite{pinson2000effect}, who conducted measurements downstream of a roughness strip placed at the leading edge, and Mushyam et al.\cite{mushyam2016numerical} who conducted numerical simulations downstream of a backward-facing step have reported Strouhal numbers of the same order. Note here that the frequency presented in the spectra in this work (across all configurations) is normalized by \textit{k} (the height of the 24-grit roughness) instead of the local displacement thickness (\textit{$\delta$*}) since we are interested in understanding how the energy associated with the shedding frequency changes with the addition of the roughness.

As we go further downstream, the flow starts to become transitional, the relative strength of the 158 Hz peak reduces, and the spectral frequency range broadens. For example, at \textit{x} = 500 mm, the spectral peak at 158 Hz has disappeared and there is an increase in the energy corresponding to frequencies both higher and lower than the 158 Hz peak. Similar observations can also be made from the pre-multiplied spectra shown in Fig. \ref{Single_strip_spectra}b. The increase in energy at higher frequencies (shown in inset in Fig. \ref{Single_strip_spectra}b) is generally attributed to the appearance of turbulent spots. \par

\begin{figure}[t]
\begin{center}
\begin{center}
    \includegraphics[width=1\textwidth]{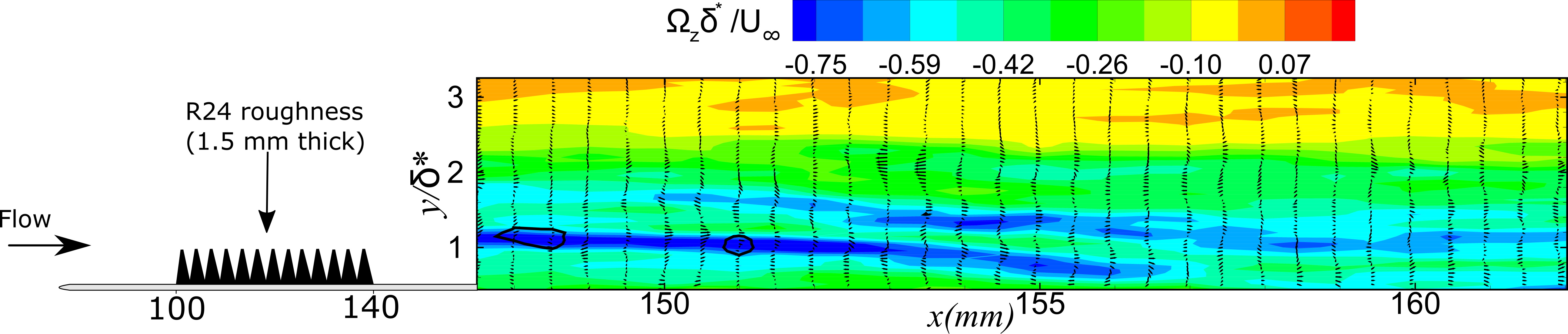}
    \end{center}
        \caption{Instantaneous spanwise vorticity contours for the single-strip configuration between \textit{x} = 145 mm and \textit{x} = 160 mm  measured using PIV. Vortices are marked using black lines. Height of the roughness approximately drawn to scale }
\label{PIV_wn_schem} 
\end{center}
\end{figure}

Particle image velocimetry measurements are taken downstream of the roughness to visualize flow features in the single-strip configuration at U$_\infty$=7.44 m/s. Figure \ref{PIV_wn_schem} shows an instantaneous frame between \textit{x} = 145 mm and \textit{x} = 165 mm in the \textit{x}-\textit{y} plane. The contours represent normalized instantaneous spanwise vorticity ($\Omega_{z}$) and spanwise vortices (denoted by black lines) are identified using the $\lambda$-criterion outlined in\cite{adrian2000analysis}. From the two-dimensional PIV data, the imaginary part of the complete eigenvalue of the local velocity gradient tensor, $\lambda$ is calculated, after which only vortices larger than the PIV interrogation size are considered. An advantage of this method is that the vortices are identified independent of their convection velocity. Around the height of the roughness (k$\sim$1.5 mm, y/$\delta^*\sim$1 at \textit{x}=155 mm), an unsteady shear layer with strong instantaneous spanwise vorticity is observed. Spanwise vortices are shed into the boundary layer around the height of the roughness and they appear in the shear layer. Choudari and Fischer\cite{choudhari2006roughness}, using spanwise vorticity contours (like the ones plotted in figure 7) had also demonstrated the presence of a shear layer and unsteady vortices being shed downstream of an array of roughness elements, which provides support to the present observation. Note that the formation of a strong shear layer downstream of a smooth backward step is a well-established feature in the literature\cite{chen2018review}.\par

Figure 8 shows instantaneous spanwise PIV measurements  between \textit{x} = 160 mm and \textit{x} = 240 mm taken inside the boundary layer at  y$\sim$1 mm (y/$\delta$* $\approx$ 0.6 at \textit{x}= 200 mm). Here, we observe the presence of streaks, which results in spanwise variation of streamwise velocity ($\partial U$ /$\partial z$). Note that the streaks are steady and this is demonstrated in \cite{joseph2022characterization}. The presence of both ($\omega_z$) and ($\partial U$ /$\partial z$) suggests that streamwise vortices can be generated by vortex tilting. Note here that streamwise vortices were found to be 'optimal' disturbances in a flat plate boundary layer that results in maximum 'transient growth' (\cite{andersson1999optimal}, \cite{luchini2000reynolds}) and in a recent study on isolated roughness induced transition, \cite{suryanarayanan2019roughness} demonstrated that streamwise vortices play a key role in streak amplification via the lift-up effect (\cite{landahl_1980}) and subsequent breakdown. Based on this, we expect that the mechanism of distributed roughness-induced transition involves streak instability (A more detailed discussion on the transition mechanism in distributed roughness is presented in \cite{joseph2022characterization}). Figure \ref{Transition steps} shows the various expected steps in distributed roughness-induced transition, based on the discussion above. The distributed roughness introduces streaks and spanwise vortices into the boundary layer. Streamwise vortices are then generated by the tilting of spanwise vortices (dashed red lines), which further results in streak amplification and transition to turbulence. Next, we investigate the flow downstream of the two-strip and three-strip configurations to quantify the transition delay and identify key factors in the transition delay phenomena.

\begin{figure}
\begin{center}
        \includegraphics[width=0.5\textwidth]{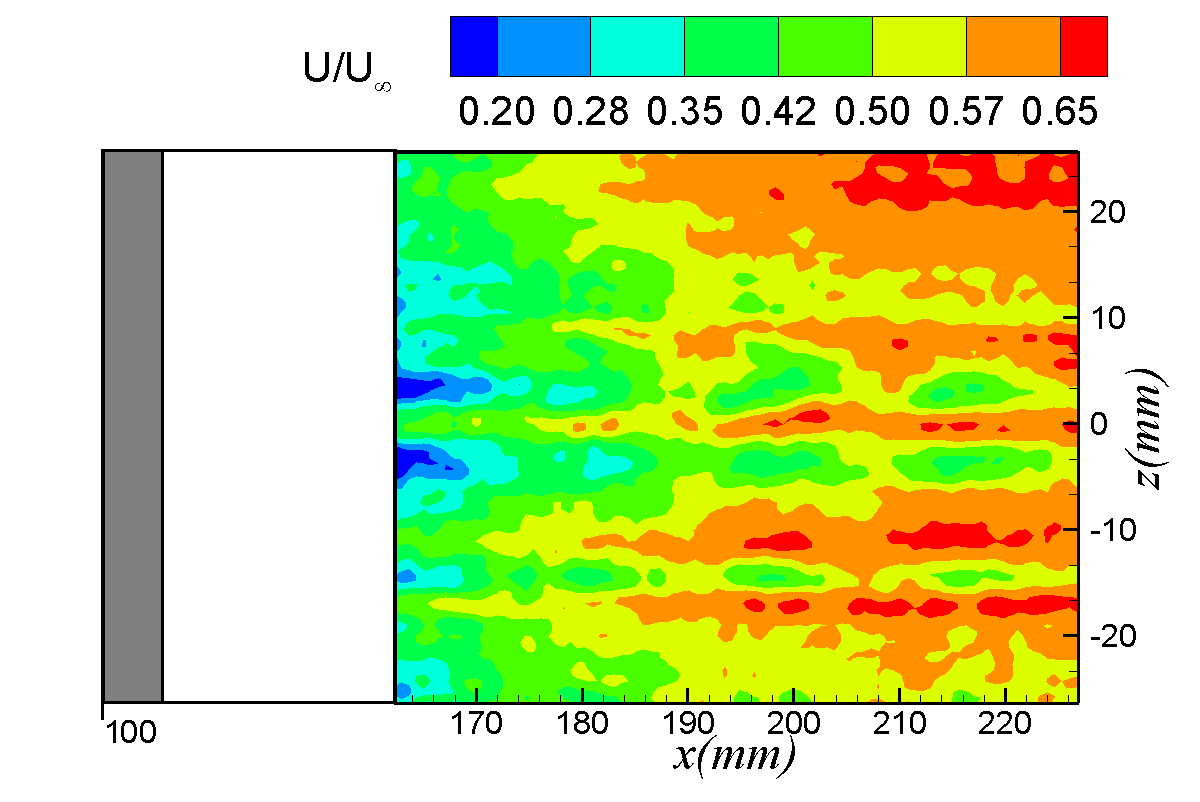}
        \caption{Instantaneous spanwise PIV measurements downstream of the distributed roughness showing the presence of streaks. The black strip denotes the distributed roughness}
         \label{Roughness_disturbance}
\end{center}
\end{figure}

\begin{figure}[t]
\begin{center}
    \includegraphics[width=1\textwidth]{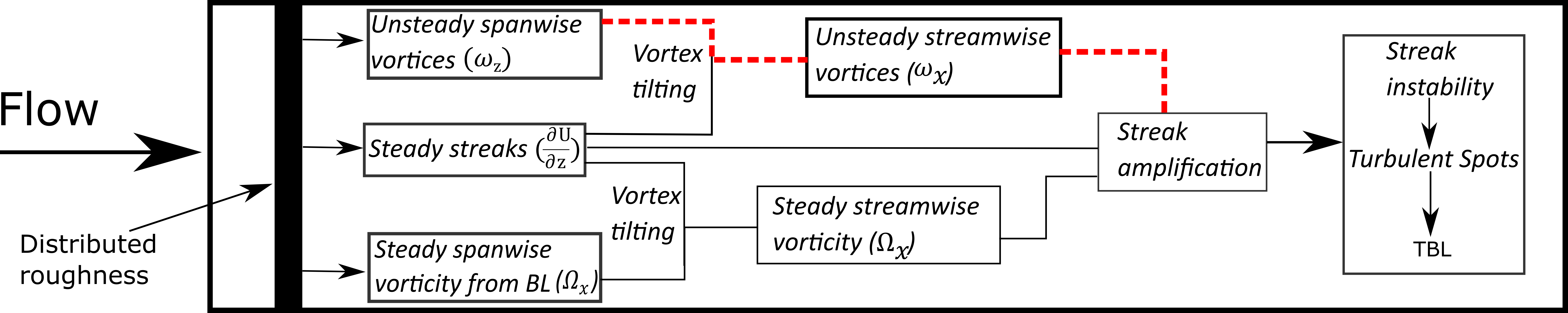}
        \caption{Expected transition mechanism for distributed roughness}
\label{Transition steps} 
\end{center}
\end{figure}

\subsubsection{Two-strip configuration}
\begin{figure}
    \centering
    \begin{minipage}{0.45\textwidth}
        \centering
        \includegraphics[width=1\textwidth]{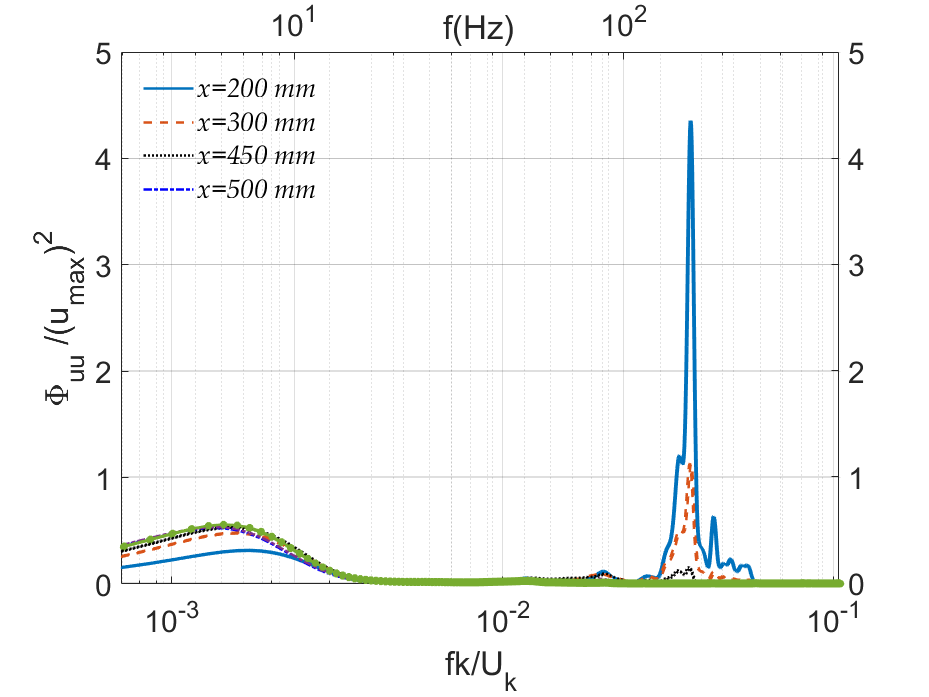} 
        \caption{Pre-multiplied spectra at different streamwise locations for two-strip configuration at U$_{\infty}$ = 7.44 m/s}
         \label{two_strip_laminar_spectra}
    \end{minipage}\hfill
    \begin{minipage}{0.45\textwidth}
        \centering
        \includegraphics[width=1\textwidth]{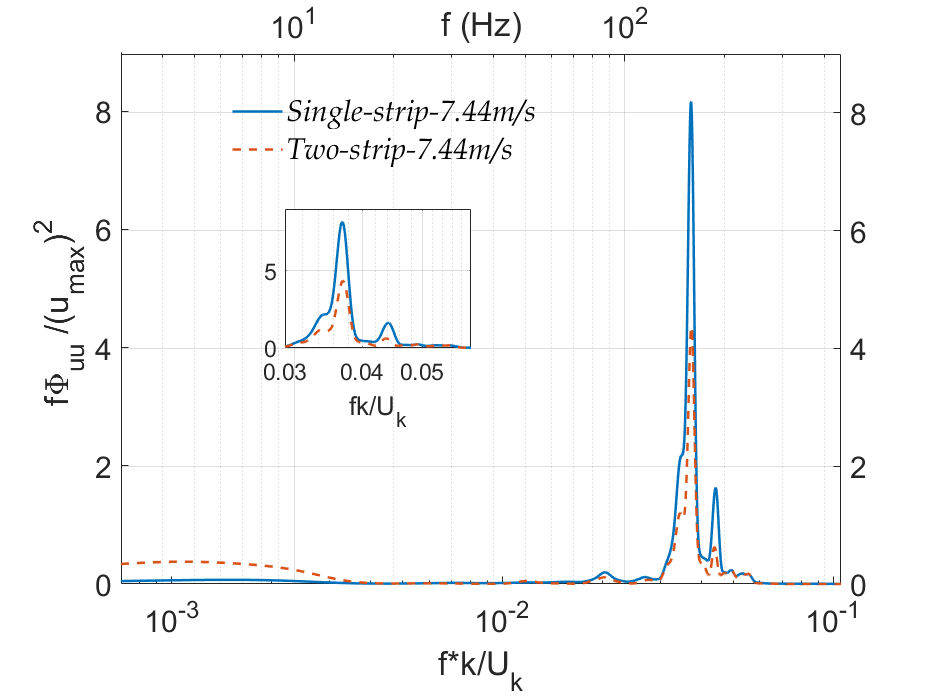} 
        \caption{Pre-multiplied spectra at \textit{x} = 200 mm for the single-strip and two-strip configurations for U$_{\infty}$ = 7.44 m/s}
        \label{single_strip_v_two_strip}
    \end{minipage}
\end{figure}

 Figure \ref{two_strip_laminar_spectra} plots the pre-multiplied spectra at different streamwise locations for two-strip configuration corresponding to the free-stream velocity, U$_\infty$=7.44 m/s and Fig. \ref{single_strip_v_two_strip} compares the pre-multiplied spectra at \textit{x} = 200 mm for the single-strip and two-strip configurations with U$_\infty$=7.44 m/s. Comparing Fig. \ref{two_strip_laminar_spectra} and Fig. \ref{single_strip_v_two_strip}, the qualitative nature of the spectra immediately downstream of both roughness configurations (i.e. at \textit{x}=200 mm) remains similar, however, the relative strength of the peak at St=0.037 has reduced considerably; for example, the normalized amplitude is reduced from approximately 8 to 4 (see Fig. \ref{single_strip_v_two_strip}). Note that the reduction in amplitude is with respect to u$_{max}$. Figure \ref{u_max} shows that u$_{max}$ for two-strip is smaller (by a half) than that for one-strip at U$_{\infty}$=7.44 m/s. Thus the absolute magnitude of the spectral peak is much smaller for the two-strip configuration as compared to single-strip. This suggests that the secondary strip downstream reduces the strength of vortices shed from the primary roughness and this could be one reason for transition delay. This observation is in agreement with that of Pinson and Wang\cite{pinson2000effect}, where they reported spectral energy concentration in a narrow wavenumber range for a single-strip configuration, but not for the case where the entire downstream region was covered with a secondary, finer roughness. However, in the present work, unlike the experiments from Pinson and Wang\cite{pinson2000effect}, transition delay was achieved using a fine roughness strip of limited streamwise extent. The reduction of spectral peak amplitude (due to the addition of secondary roughness strip downstream) correlating to transition delay suggests that there is lesser amount of spanwise vorticity available to be tilted into streamwise vorticity via the vortex tilting mechanism (red dashed lines in figure \ref{Transition steps}). Due to this, the overall amount of streamwise vorticity generated is lower in the two-strip configuration (in comparison to the single-strip configuration) and this results in transition delay. \\
Figure \ref{two_strip_trans_spectra} plots the pre-multiplied spectra for the two-strip configuration undergoing transition at U$_\infty$=8.47 m/s. Comparing with Fig. \ref{two_strip_laminar_spectra}, higher frequency peaks appear in the spectra pointing to new length scales being introduced in the flow as the two-strip configuration undergoes transition. The downstream development of the spectra is similar to the single-strip configuration undergoing transition with the higher frequency peaks dying out and spectral energy in the low frequency range increasing as the flow becomes transitional.\\

\subsubsection{Three-strip configuration}
\begin{figure}
    \centering
    \begin{minipage}{0.45\textwidth}
        \centering
        \includegraphics[width=1\textwidth]{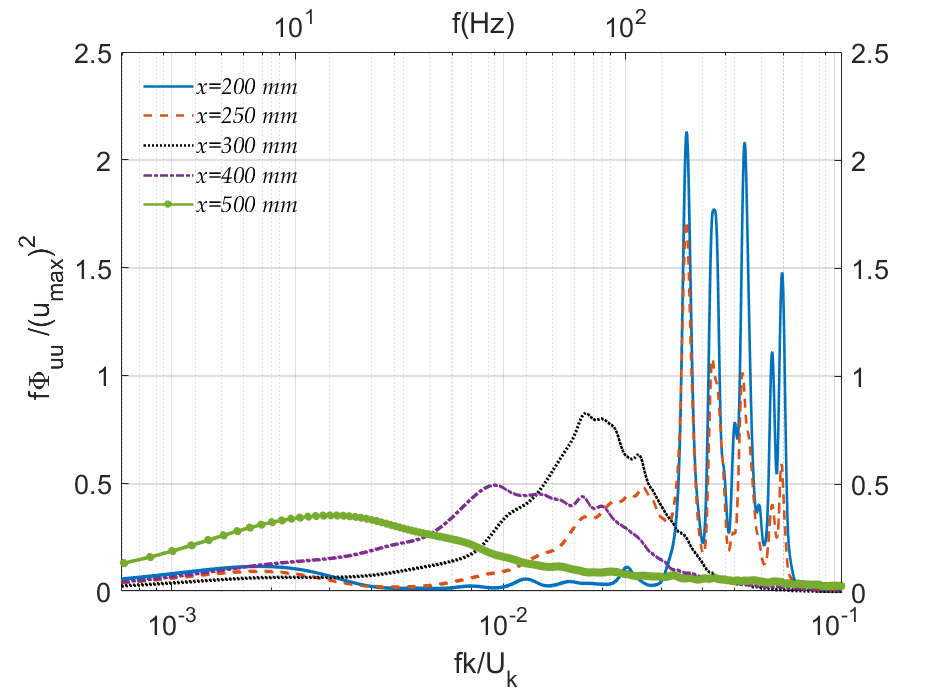} 
        \caption{Pre-multiplied spectra at different streamwise locations for two-strip configuration at  U$_\infty$=8.47 m/s}
         \label{two_strip_trans_spectra}
    \end{minipage}\hfill
    \begin{minipage}{0.45\textwidth}
        \centering
        \includegraphics[width=1\textwidth]{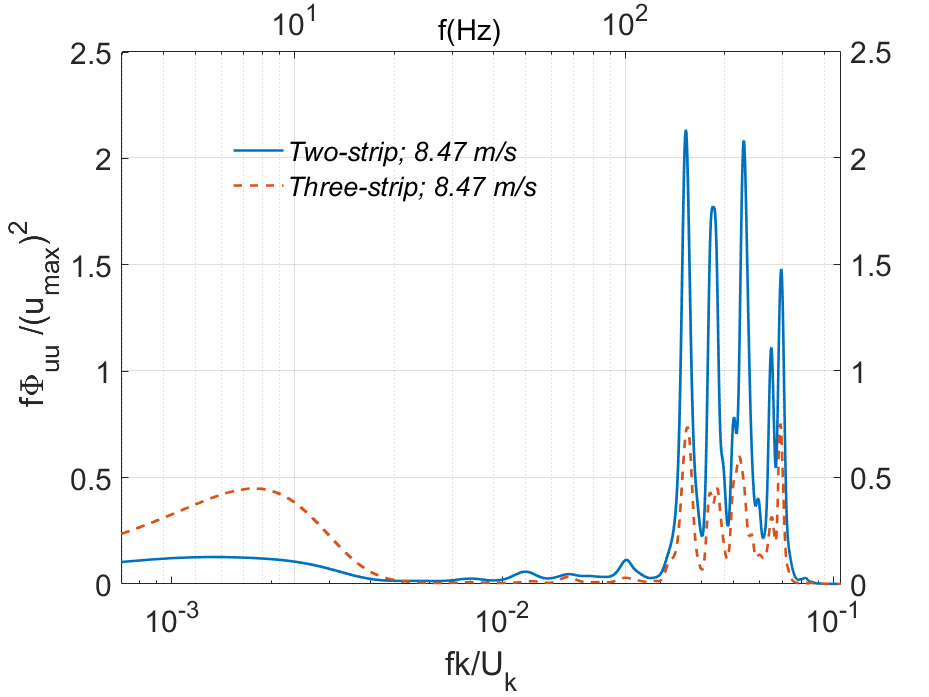} 
        \caption{Pre-multiplied spectra at \textit{x} = 200 mm for two-strip configuration and three-strip configuration at    U$_\infty$=8.47 m/s}
        \label{two_strip_v_three_strip_spectra}
    \end{minipage}
\end{figure}

 Figure\ref{two_strip_v_three_strip_spectra} compares the velocity spectra for two-strip and three-strip configurations at \textit{x} = 200 mm and  U$_{\infty}$=8.47 m/s and here too(like Fig. \ref{single_strip_v_two_strip}), we see that the relative strength of the peaks reduces with the addition of the upstream roughness strip.
The pre-multiplied spectra at different streamwise locations for the three-strip configuration undergoing transition corresponding to free-stream velocity of 8.81 m/s is shown in Fig. \ref{Three_strip_trans_spectra}. The high-frequency peaks dying out and spectral energy increasing in the lower frequency region(as we go downstream) is observed here as well. However, unlike the spectra of the two-strip configuration at U$_{\infty}$=8.47 m/s (Fig. \ref{two_strip_trans_spectra}), in which new high frequency peaks were observed as the flow becomes transitional, in the three-strip configuration undergoing transition(at U$_{\infty}$=8.81 m/s), we do not see new length scales/frequencies introduced into the boundary layer. This is likely because the increase in velocity required for the flow to become transitional is higher between the single-strip and two-strip configuration(7.44 m/s to 8.47 m/s; 13\% increase) in comparison to two-strip and three-strip configuration(8.47 m/s to 8.81 m/s; 4\% increase)\\

\subsection{Mechanism of transition delay by secondary roughness}

In this section, we discuss the possible mechanisms by which the secondary fine roughness strips placed upstream/downstream delays transition. Previous studies have reported delaying boundary layer transition caused by a discrete roughness element by placing a distributed roughness strip downstream. Further,  Pinson and Wang\cite{pinson2000effect} showed that transition due to distributed surface roughness can be delayed by covering the entire downstream region with finer grit roughness. They report that covering the downstream region with finer grit roughness will reduce the strength of vortices shed from the primary roughness. In the present work, the secondary roughness strip placed downstream reduces the strength of spanwise vortices shed from the primary roughness due to a reduction in step size. This is supported by the  spectral analysis of hotwire measurements, where we saw a decrease in the relative strength of spectral peaks for the two-strip configuration in comparison to the single-strip configuration. The reduction of spanwise vorticity generated from the primary roughness means that there is lesser spanwise vorticity available to be tilted into streamwise vorticity. The reduction in streamwise vorticity, which is expected to play a key role in distributed roughness-induced transition (figure \ref{Transition steps}, see also \cite{drews2012direct}) results in transition delay. Note that it is also possible for the secondary roughness downstream to cause dissipation of streamwise vorticity generated by the primary roughness (as discussed in \cite{lu2020investigation}, \cite{suryanarayanan2020mechanisms}, \cite{suryanarayananmechanics})\par

\begin{figure}
    \centering
    \begin{minipage}{0.45\textwidth}
        \centering
        \includegraphics[width=1\textwidth]{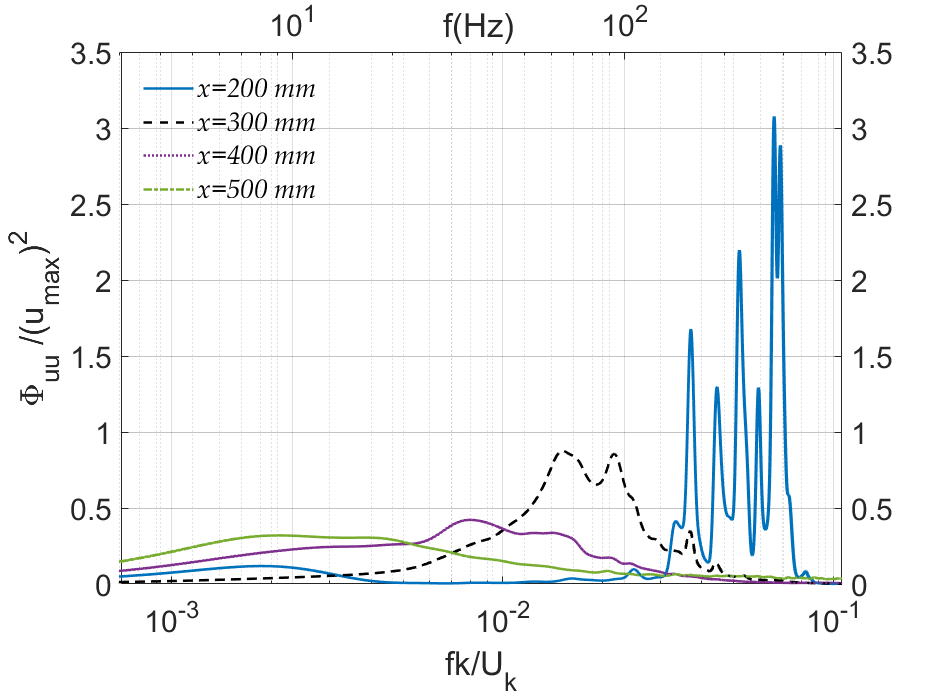} 
        \caption{Pre-multiplied spectra at different streamwise locations for three-strip configuration at  U$_\infty$=8.81 m/s}
         \label{Three_strip_trans_spectra}
    \end{minipage}\hfill
    \begin{minipage}{0.45\textwidth}
        \centering
        \includegraphics[width=1\textwidth]{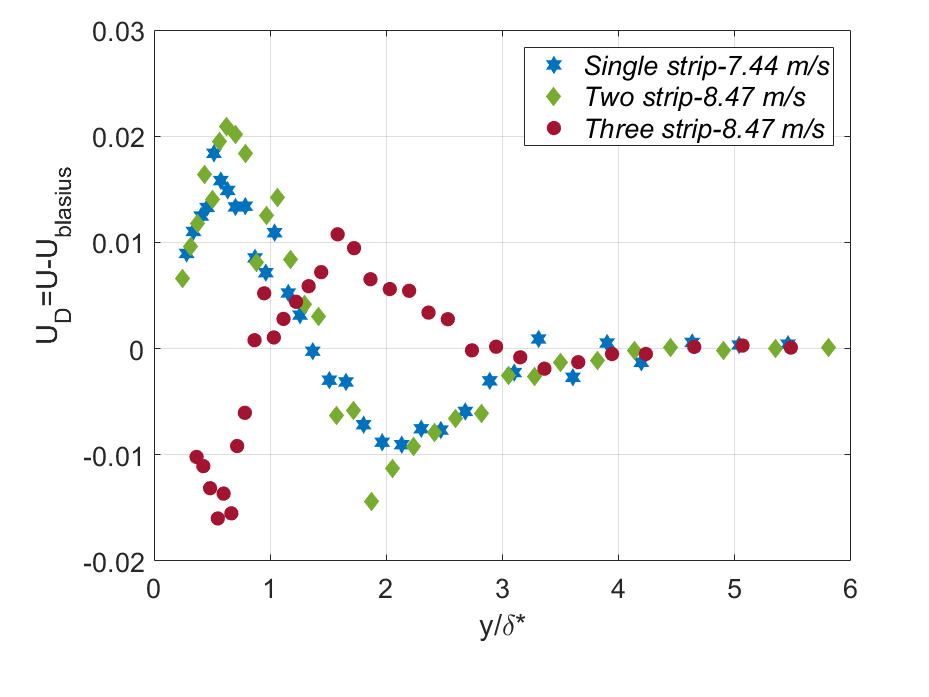}
        \caption{Mean disturbance profiles at \textit{x}=200 mm for different configurations}
        \label{Mean_distur}
    \end{minipage}
\end{figure}

Previous numerical studies by Suryanarayanan et al.\cite{suryanarayanan2020mechanisms} demonstrated delaying transition of an isolated roughness element using a distributed roughness or flat strip placed upstream. The transition delay is due to lifting of the boundary layer by the upward strip, thereby reducing the effective Re$_k$ of the isolated roughness element. We expect that the roughness strip placed upstream in the three-strip configuration in the present measurements will also have a similar effect. 
The lifting of the boundary layer can be quantified by plotting the deviation of mean velocity from the Blasius profile\cite{kuester2015roughness}.  Near the wall, if the mean velocity is lower than the Blasius velocity, it means that the boundary layer is lifted because of the presence of the upstream roughness\cite{kuester2015roughness}. It is expected that if the upstream roughness lifts the boundary layer, that effect might be present not just near the roughness, but also some distance downstream at \textit{x} = 200 mm. Figure\ref{Mean_distur} plots the mean disturbance profiles at \textit{x} = 200 mm for the two- and three-strip configuration at U$_\infty$=8.47 m/s along with single-strip configuration at U$_\infty$=7.44 m/s. In case of the two-strip configuration(which has secondary roughness only downstream) at U$_\infty$=8.47 m/s and single-strip configuration at U$_\infty$=7.44 m/s, the velocity profile at \textit{x} = 200 mm is pre-transitional. The mean disturbance profile being positive close to the wall and then becoming negative as we move away from the wall is a feature of pre-transitional flow and has been reported in FST induced transition as well\cite{matsubara2001disturbance}. In comparison, in the three-strip configuration at U$_\infty$=8.47 m/s, the flow is laminar but the mean velocity close to the wall is less than the Blasius value. In concurrence with Kuester and White\cite{kuester2015roughness}, we attribute this to the lifting of the boundary layer by the upstream roughness, the effect of which persists downstream and is seen at \textit{x} = 200 mm in Fig. \ref{Mean_distur}. Because of this lifting of the boundary layer by the upstream secondary roughness in the three-strip configuration, the effective Re$_k$ of the primary roughness will decrease. This could be the reason why the flow is laminar (Fig. \ref{u_inst}d) in case of three-strip configuration, while flow is transitional in two-step configuration (Fig. \ref{u_inst}c) at \textit{x} = 500 mm. Another possible mechanism is that the presence of the upstream secondary roughness will weaken the separation bubble created by the forward step of primary roughness. This will reduce the strength of disturbances created by the bubble, thus lessening their effect on transition. 

In the three-strip configuration, there is the combined effect of upstream roughness strip and downstream roughness strip delaying transition, due to which it shows higher transition delay than the two-strip configuration.

\subsection{Effect of length and type of the downstream secondary roughness}

To ensure the repeatability of transition delay caused by the downstream roughness strips and to investigate the effect of changing the length of the downstream roughness, an additional set of measurements was conducted in a separate wind tunnel with a 60 cm x 60 cm cross-section and freestream turbulence intensity $\sim$0.1\%. The measurements were conducted on a flat plate with a sharp leading edge and the primary roughness used was a 24-grit sandpaper strip with 40 mm streamwise extent placed at 100 mm from the leading edge.  

\begin{figure}[t]
\begin{center}
    \includegraphics[width=0.6\textwidth]{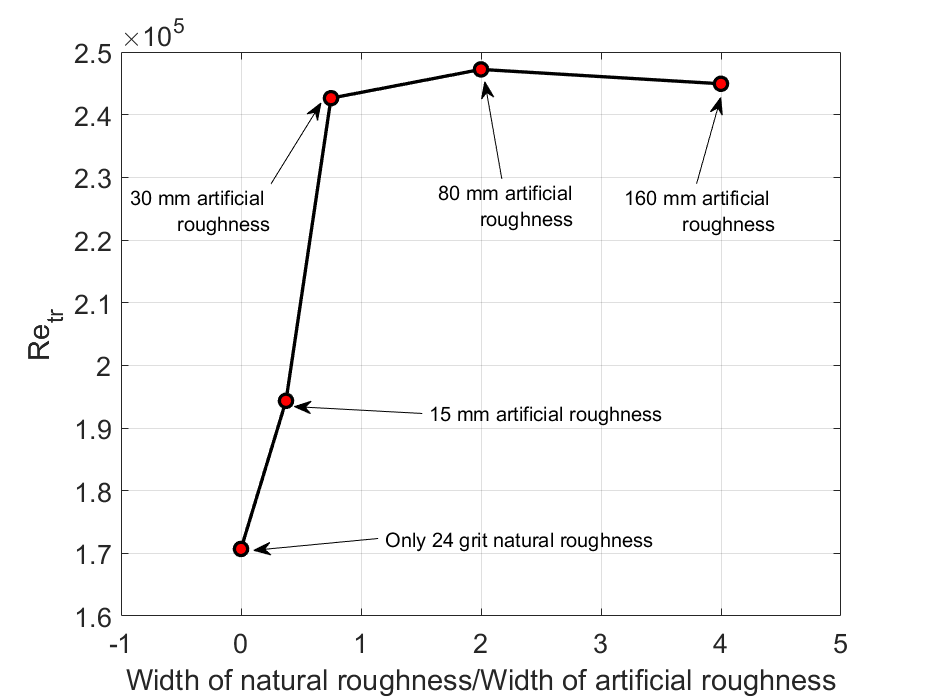}
        \caption{Streamwise Reynolds number (Re$_{x}$) corresponding to the appearance of turbulent spots for different downstream artificial roughness lengths}
\label{Trans_onset_length} 
\end{center}
\end{figure}

For this set of measurements, we are interested in analyzing the qualitative change in transition behavior as the length of the secondary roughness strip is increased. Towards this, we monitor the freestream velocity for the appearance of turbulent spots for various configurations, consistent with \cite{hernon2007experimental}. The effect of increasing the length of the downstream secondary strip on the streamwise Reynolds number (Re$_{x}$=U$_{\infty}x/\nu$, \textit{x} is the streamwise coordinate) for the appearance of turbulent spots is plotted in figure  \ref{Trans_onset_length}. Note that based on prior observations, the streamwise Reynolds number at transition onset (based on  $\sim$ 10\% intermittency) is approximately 10\% higher than the streamwise Reynolds number at the location where turbulent spots just start to appear. Here, we notice that increasing the length of the downstream secondary roughness from 15 mm to 30 mm delayed the appearance of turbulent spots from a Reynolds number of 1.93x10$^5$ to 2.40x10$^5$, however increasing the length of the downstream roughness beyond 30 mm did not substantially affect transition delay. It is relevant to discuss this result in the context of \cite{pinson2000effect} who observed transition delay using a similar arrangement. In that study, they covered the \textit{entire} downstream region with a secondary roughness and observed transition delay. The present results suggest that there maybe an optimum length of the secondary downstream roughness (that could vary based on primary roughness and the secondary roughness used) that results in maximum transition delay, beyond which, increasing the length of the secondary downstream roughness is unlikely to improve performance.\par

\begin{figure}[t]
\begin{center}
    \includegraphics[width=1\textwidth]{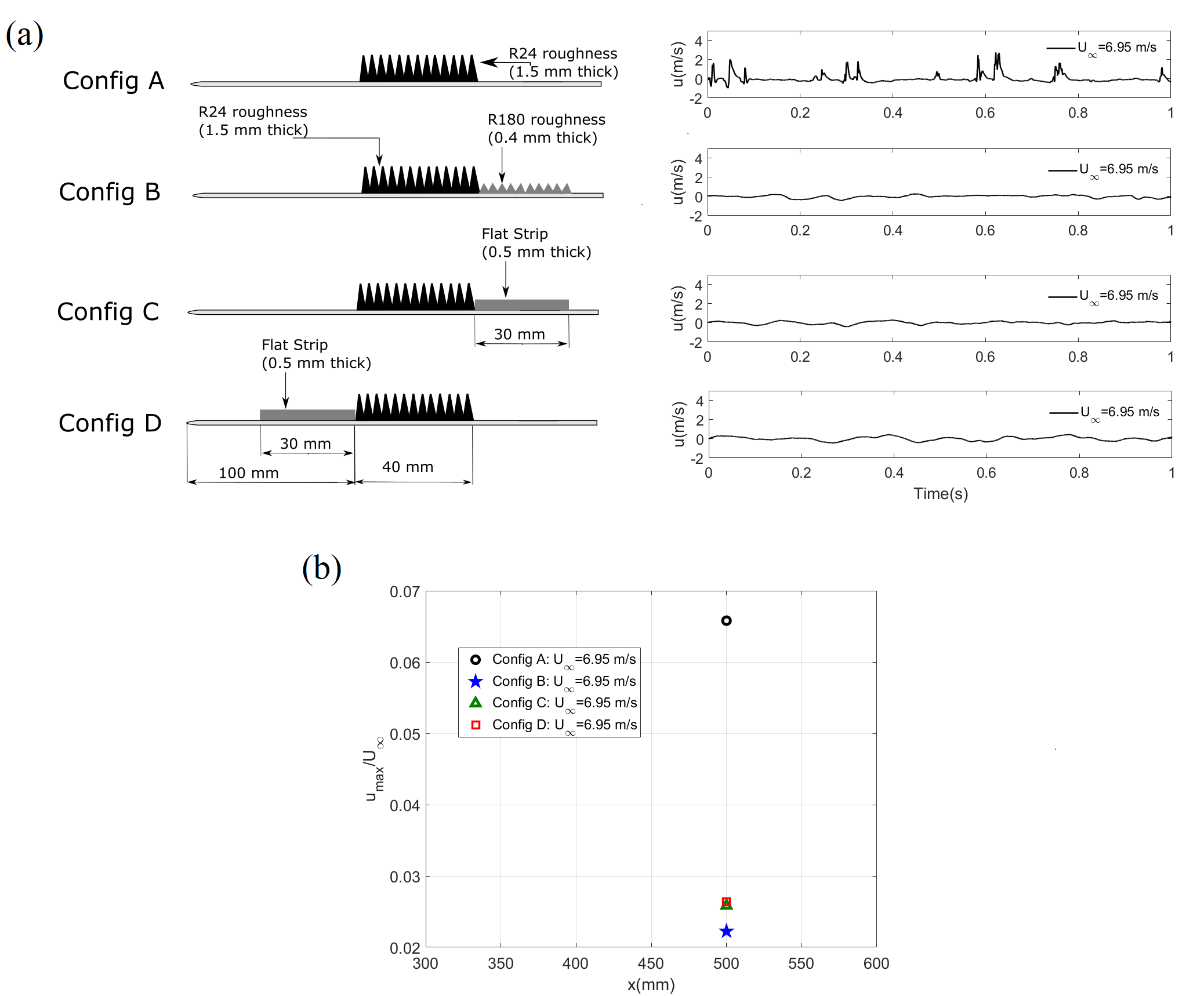}
        \caption{(a) Demonstration of transition delay using different types of roughness strips (b)Decrease in maximum rms due to roughness}
\label{roughness_para} 
\end{center}
\end{figure}

Finally, we investigate whether a smooth strip can act as a secondary strip to delay transition. Toward this, we conduct measurements for four different roughness configurations (Config A-D) at \textit{x}=500 mm for the same freestream velocity (figure\ref{roughness_para}). Config A contains only the 24-grit roughness, while Config B contains a 180-grit secondary roughness strip placed downstream, Config C contains a smooth strip placed downstream and Config D contains a smooth strip placed upstream. Here, we see that at \textit{x}=500 mm, Config A shows causes turbulent spots, while in the other three configurations, the flow was laminar, as evidenced by the lack of turbulent spots in the velocity signal in figure\ref{roughness_para}a and a corresponding reduction in maximum rms levels in figure\ref{roughness_para}b. This demonstrates that a smooth strip can also function as a secondary strip to delay boundary layer transition caused by naturally occurring roughness. It is pertinent to note here that transition delay using a smooth strip placed downstream has been demonstrated numerically by Lu et al. \cite{lu2020investigation} and Suryanarayanan et al.\cite{suryanarayananmechanics} for boundary layer transition caused by isolated and distributed roughness respectively.\par

This technique of using downstream roughness of limited length can have several applications in practical engineering scenarios like wind turbine blades, aircraft wings, and gas turbine blades where the secondary roughness strip can be pre-applied in scenarios where we expect roughness accumulation: for example, an aircraft/UAV about to fly in icing conditions, or a wind turbine blade exposed to dust or insects. This is currently under patent application.

\begin{figure}[t]
\begin{center}
    \includegraphics[width=1\textwidth]{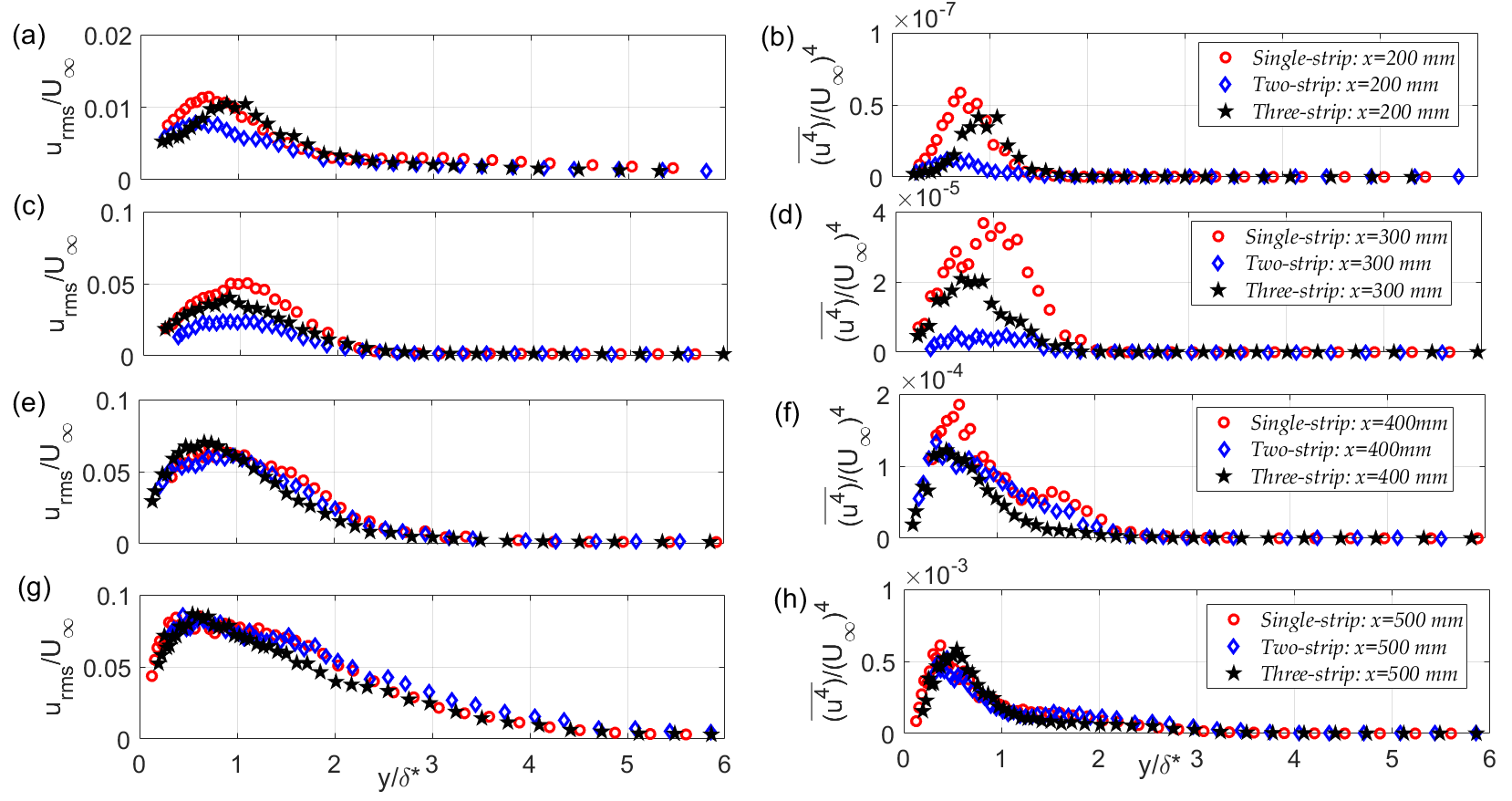}
        \caption{Comparison of second (u$_{rms}$) and fourth moments ($\overline{u}\ ^4$) of fluctuating velocity normalized with freestream velocity at different streamwise locations for each configuration undergoing transition}
\label{trans_combined} 
\end{center}
\end{figure}

\begin{figure}[t]
\begin{center}
    \includegraphics[width=0.5\textwidth]{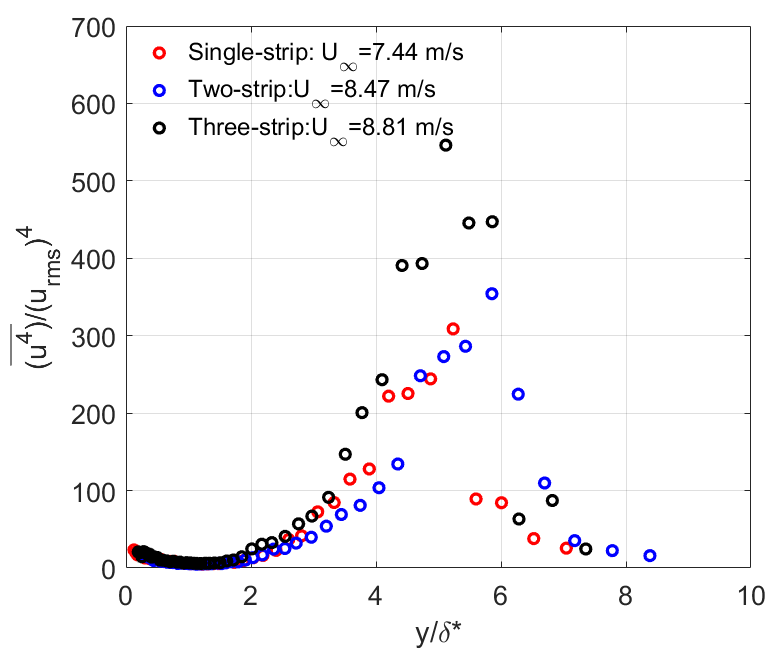}
        \caption{Wall-normal variation of flatness at \textit{x} = 500 mm at transitional flow velocities for different configurations}
\label{Re_transition} 
\end{center}
\end{figure}

\subsection{Effect of secondary roughness on transitional flow}

While the present study demonstrates that secondary roughness strips placed upstream or downstream can delay boundary layer transition caused by primary roughness, it is of interest to see whether the secondary roughness introduces any specific signatures into the  transitional flow in each of these configurations. Note that we already observed from comparing the spectra in the pre-transitional region that the secondary roughness introduces additional peaks (comparing figures 12 and 13). Next, we compare the wall-normal profiles of second and fourth velocity moments at different streamwise locations of the three roughness configurations. Note here that the fourth moment is normalized with the freestream velocity and not root-mean-square (rms) of fluctuating velocity (as is typically done while calculating skewness). This is because normalizing with the RMS of fluctuating velocity will lead to very high values outside the boundary layer, primarily due to the low fluctuation levels in the freestream (see figure \ref{Re_transition}). 

From figure \ref{trans_combined}, it is clear that the profiles corresponding to the second and fourth moments are not similar in the early stages of transition. However, after the onset of transition (figure \ref{trans_combined} g and h), the profiles collapse fairly well. From section 3.1 and figure 12, we already know that the maximum rms levels and intermittency are comparable for the three configurations after the onset of transition. This indicates that \textit{after the onset of transition}, there aren't any discernible features in the flow corresponding to the addition of the secondary roughnesses: i.e. the transitional flow can be modeled as if it was caused by a single roughness. This points to another possible advantage of the transition delay mechanism using additional strips of distributed roughness: in addition to being lightweight and easy to implement, the secondary strips (based on the present study) do not seem to have introduced any specific signatures into the flow after the onset of transition.

\section{Conclusions}
A passive method for delaying distributed roughness-induced transition using fine roughness strips placed upstream or downstream of the primary roughness is presented. Results show that a fine-grit distributed roughness strip (secondary roughness) placed downstream of a primary roughness can delay boundary layer transition caused by the primary roughness. A three-strip configuration consisting of roughness strips placed upstream and downstream is most effective in delaying transition and increases the free stream velocity required for the onset of transition by $\sim$18\% in comparison to the single-strip configuration. A parametric study shows that increasing the length of the downstream secondary roughness strip increases the transition delay, however, there is likely an optimal length of the secondary roughness beyond which increasing the length does not meaningfully increase the transition delay. Further, it is also demonstrated that a smooth strip, placed upstream or downstream can also be used as a secondary strip to delay boundary layer transition.

PIV and hot-wire measurements downstream of the roughness demonstrate that the primary roughness sheds spanwise vortices into the boundary layer while also causing steady streaks. Analysis of the hot-wire signals suggests that the mechanism by which upstream and downstream roughness delays transition are different.  The downstream roughness weakens the strength of spanwise vortices shed from the primary roughness as evidenced by the reduction in strength of the spectral peak. This reduces the strength of streamwise vortices generated via vortex tilting, thereby weakening a key step in the expected transition mechanism. The upstream roughness delays transition by ‘lifting’ the boundary layer and reducing the effective roughness Reynolds number of the primary roughness and a supporting argument is made by analyzing the mean disturbance profiles downstream. Finally, it is demonstrated that while different roughness configurations have their own signatures (in the form of spectral peaks) in the pre-transitional flow, after the onset of transition, based on the higher-order velocity moments, the flow characteristics of all three roughness configurations are similar. This points to a potential utility of the roughness configurations, wherein the additional roughness strips do not show any specific signatures into the flow after the onset of transition.  

The present work points to a lightweight, inexpensive, and easy-to-implement method to delay the boundary layer transition caused by random distributed roughness accumulating on aerodynamic surfaces. The results demonstrate the effect of various types of distributed roughness and provides insight into the mechanism of transition delay.

\bibliography{main}

\end{document}